\documentclass[printer,oldversion]{aa}
\usepackage{natbib}
\usepackage{graphicx}
\usepackage{color}
\usepackage{txfonts}

\bibpunct{(}{)}{;}{a}{}{,}

\begin{document}

\title{Full-Sky Weak-Lensing Simulation with 70 Billion Particles}

\author{Romain Teyssier \inst{1,4} \and Sandrine Pires\inst{1,2}
 \and Simon Prunet\inst{4} 
 \and Dominique Aubert\inst{3} \and Christophe Pichon\inst{1,4}  
 \and Adam Amara\inst{1}  \and Karim Benabed\inst{4} 
 \and St\'ephane Colombi\inst{4} \and Alexandre Refregier\inst{1} 
 \and Jean-Luc Starck\inst{1,2}}

\offprints{romain.teyssier@cea.fr}

\institute{Service d'Astrophysique,
CEA Saclay, Batiment 709, 91191 Gif--sur--Yvette Cedex, France \and
Service d'Electronique, de Detection et d'Informatique,
CEA Saclay, Batiment 141, 91191 Gif--sur--Yvette Cedex, France \and
Observatoire Astronomique, Universit\'e de Strasbourg, UMR 7550, 
11 rue de l'Universit\'e, 67000, Strasbourg, France \and
Institut d'Astrophysique de Paris, 
98 bis, boulevard Arago, 75014 Paris, France}
\date{Accepted; Received; in original form;}

\abstract{We have performed a 70 billion dark-matter particles N--body
  simulation in a 2 $h^{-1}$  Gpc periodic box, using the concordance,
  cosmological model as  favored by the latest WMAP3  results. We have
  computed  a full-sky convergence  map with  a resolution  of $\Delta
  \theta \simeq 0.74$ arcmin$^{2}$,  spanning 4 orders of magnitude in
  angular dynamical  range.  Using various high-order  statistics on a
  realistic  cut sky, we  have characterized  the transition  from the
  linear to the nonlinear regime  at $\ell \simeq 1000$ and shown that
  realistic  galactic masking affects  high--order moments  only below
  $\ell  < 200$.   Each domain  (Gaussian and  non--Gaussian)  spans 2
  decades in angular  scale.  This map is therefore  an ideal tool for
  testing map--making  algorithms on  the sphere. As  a first  step in
  addressing the full map  reconstruction problem, we have benchmarked
  in this paper two denoising  methods: 1) Wiener filtering applied to
  the  Spherical Harmonics  decomposition  of  the map  and  2) a  new
  method,  called MRLens,  based on  the modification  of  the Maximum
  Entropy  Method on  a Wavelet  decomposition.  While  the  latter is
  optimal  on large  spatial  scales, where  the  signal is  Gaussian,
  MRLens outperforms the Wiener  method on small spatial scales, where
  the  signal is  highly  non--Gaussian.  The  simulated full-sky
  convergence map  is freely  available to the  community to  help the
  development  of new  map--making  algorithms dedicated  to the  next
  generation of weak--lensing surveys.}

\keywords{Methods: N--body simulations; Cosmology: observations; 
Techniques: image processing}

\authorrunning{Teyssier {\it et al. }}
 
\titlerunning{Full-sky weak-lensing simulation with 70 billion particles}

\maketitle

\section{Introduction}

Weak  gravitational lensing,  or ``cosmic  shear'', provides  a unique
tool for mapping the matter  density distribution in the Universe (for
reviews,     see     \cite{refregier03},     \cite{hoekstra03}     and
\cite{munshi06}).  Current weak-lensing surveys cover altogether about
100 square degrees and have been  used to measure the amplitude of the
matter   power  spectrum  and   other  cosmological   parameters  (see
\cite{benjamin07}, \cite{fu08}  and references therein).   A number of
new  instruments are  being planned  to carry  out these  surveys over
wider  sky areas (PanSTARRS,  DES, SNAP  and LSST)\footnote{PanSTARRS:
  http://pan-starrs.ifa.hawaii.edu,                                DES:
  https://www.darkenergysurvey.org,   SNAP:   http://snap.lbl.gov  and
  LSST: http://www.lsst.org}  or even over the  full extragalactic sky
(DUNE\footnote{DUNE:  http://www.dune-mission.net}).  These wide-field
surveys  will yield  cosmic-shear measurements  on both  large scales,
where  gravitational  dynamics is  in  the  linear  regime, and  small
scales,  where the dynamics  is highly  nonlinear.  The  comparison of
these measurements  with theoretical predictions of  the density field
evolution  will place strong  constraints on  cosmological parameters,
including dark energy  parameters (e.g. \cite{hu99}, \cite{huterer01},
\cite{amara06}  and \cite{albrecht07}).  On  small scales,  the highly
nonlinear nature  of the density field ensures  that predictions based
on analytic calculations are  prohibitively difficult and requires the
use of numerical simulations.   N-body simulations have thus been used
to simulate weak-lensing  maps across small patches of  the sky, using
the  flat sky approximation  (e.g. \cite{jain00},  \cite{hamana01} and
\cite{vale04}).  The  simulation of  full-sky maps in  preparation for
future surveys involve a wide range of both mass and length scales and
is challenging  for current N-body  simulations.  The range  of scales
involved  also requires  the development  of efficient  algorithms for
deriving a mass map from  true noisy data sets.  These algorithms need
to  be  well--suited  to   both  the  large--scale  signal,  which  is
essentially a Gaussian random field, and those on small--scales, where
it is highly non--Gaussian and exhibits localized features.

\begin{figure*}
\centering
\resizebox*{!}{9cm}{\includegraphics{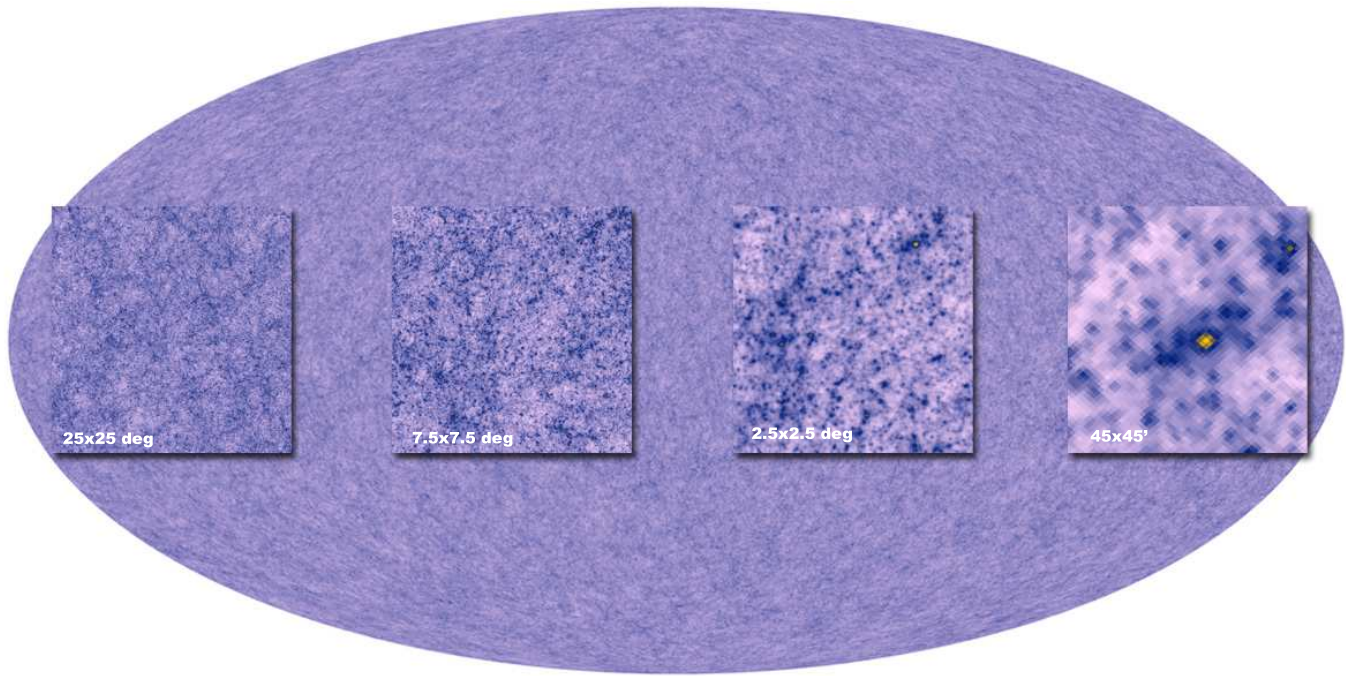}}
\vskip -0.1cm
\caption{Full-sky simulated convergence map derived from the {Horizon}
  Simulation.   Its   resolution  of  200  million   pixels  has  been
  downgraded  to fit  the page.   The various  inserts display  a zoom
  sequence into smaller and smaller  areas of the sky.  The pixel size
  is 0.74 arcmin$^{2}$.  }
\label{fullsky}
\end{figure*}

In  this  paper,  we  used  a high  resolution  N-body  simulation  to
construct   a    full-sky   weak-lensing   map   and    test   a   new
map-reconstruction method based  on a multi-resolution technique.  For
this purpose,  we use  the Horizon simulation,  a 70  billion particle
N--body simulation, featuring more than  $140$ billion cell in the AMR
grid of  the RAMSES code \citep{teyssier02}.  The  simulation covers a
sufficiently  large  volume   ($L_{box}=2  h^{-1}$Gpc)  to  compute  a
full-sky convergence  map, while  resolving Milky-Way size  halos with
more than 100 particles, and exploring small scales into the nonlinear
regime  (see   Sect.~\ref{simu}).   This  unprecedented  computational
effort allows us, for the first  time, to close the gap between scales
close to  the cosmological horizon  and scales deep  inside virialized
dark-matter  haloes.   A  similar   effort  at  lower  resolution  was
presented by \cite{fosalba08}.

The  dark-matter distribution in  the simulation  was integrated  in a
light  cone to  a redshift  of 1,  around an  observer located  at the
centre of the simulation  box (see Sect.~\ref{cone}).  This light cone
was  then  used  to   calculate  the  corresponding  full-sky  lensing
convergence field, which  we mapped using the Healpix\footnote{HeaPix:
  http://healpix.jpl.nasa.gov}  pixelisation  scheme  \citep{gorski05}
with a  pixel resolution of  $\Delta \theta \simeq  0.74$ arcmin$^{2}$
($n_{\rm side}=4096$), and added  ``instrumental'' noise for a typical
all--sky  survey with 40  galaxies per  arcmin$^{2}$, as  expected for
example  for the  DUNE mission  \citep{dune06}.  Using  an Undecimated
Isotropic   Wavelet   Decomposition   of  this   realistic   simulated
weak-lensing map  on the  sphere, we analyzed  the statistics  of each
wavelet plane  using second, third and fourth  order moments estimator
(Sect.~\ref{stat}).   We   then  applied,  in   Sect.~\ref{filter},  a
multi-resolution algorithm  to filter a  fictitious simulated $\kappa$
data set based  on an extension of the  wavelet filtering technique of
\citet{pires06}.  We  characterised the quality  of the reconstruction
using the  power spectrum  of the  error map and  compare this  to the
result  of standard  Wiener  filtering on  the  sphere.  Our  results,
summarised  in Sect.~\ref{conclusion}, illustrate  the virtue  of high
resolution simulations  such as the  one reported here to  prepare for
future weak-lensing surveys and to design new map--making techniques.

\section{The Horizon N-Body simulation}
\label{simu}
 
This large  N-body simulation  was carried out  using the  RAMSES code
\citep{teyssier02} for  two months on the 6144  Itanium2 processors of
the  CEA  supercomputer  BULL  Novascale  3045  hosted  in  France  by
CCRT\footnote{Centre de Calcul Recherche et Technologie}.  RAMSES is a
parallel hydro and  N-body code based on the  Adaptive Mesh Refinement
(AMR)  techniques.  Using  a parallel  version of  the \texttt{grafic}
package \citep{bertschinger01}, we  generated the initial displacement
field on a $4096^3$ grid for the cosmological parameters from the WMAP
3rd   year    results   \citep{spergel07},   namely   $\Omega_m=0.24$,
$\Omega_\Lambda=0.76$, $\Omega_b=0.042$,  $n=0.958$, $H_0=73$ km/s/Mpc
and  $\sigma_8=0.77$.    We  used  the   \cite{eisenstein98}  transfer
function, which includes baryon oscillations.  The box size was set to
2  Gpc/h, which  corresponds roughly  to the  comoving distance  to an
object at $z  \simeq 0.8$.  We used 68.7  billion particles to simuate
the dark-matter density field,  yielding a particle mass of $7.7\times
10^9$  $M_\odot$ and  130 particles  per Milky  Way halo.   This large
particle distribution was split  across 6144 individual files, one for
each  processor, according  to  the RAMSES  code domain  decomposition
strategy \citep{aubert07}.   Starting with a base (or  coarse) grid of
$4096^3$ grid points, AMR cells  are recursively refined if the number
of  particles in  the cell  exceeds 40.   In this  way, the  number of
particles per cell varied between 5  and 40, so that the particle shot
noise remained at an acceptable  level.  At the end of the simulation,
we had reached 6 levels of  refinement with a total of 140 billion AMR
cells.  This corresponds to a  formal resolution of $262~144^3$ or 7.6
$h^{-1}$  kpc  comoving  spatial  resolution.  Parallel  computing  is
perfomed  using  the  MPI  message--passing  library,  with  a  domain
decomposition based  on the Peano--Hilbert  space--filling curve.  The
work and memory load was adjusted dynamically by reshuffling particles
and grid points from each  processor to its neighbors.  The simulation
required 737  main (or  coarse) time steps  and more than  $10^4$ fine
time steps for completion.

\section{Light cone and convergence map}
\label{cone}

Born's  unperturbed-trajectory assumption  for  all neighboring  light
rays  is  a good  approximation  in  the  linear regime  of  structure
formation, but  is inaccurate in the  nonlinear regime.  Consequently,
distortion  effects  of  lensing  beyond  the first  order  cannot  be
simulated  reliably.   As  shown  by  \cite{vanwaerbeke01},  the  Born
approximation also introduces a relative  error in the skewness of the
signal of aproximately  10\% on large scales where  the convergence is
Gaussian, and about  1\% on small scales in  the nonlinear regime.  We
therefore implemented  a multiple-lens ray-tracing method  that can be
applied more generally than Born's approximation.

We constructed a light cone by  recording, at each main time step, the
positions of particles  within the boundaries of a  photon plane: this
plane moved at the speed of light towards an observer, who was located
at  the centre  of the  box.  Our  method was  developed from  the one
presented by \cite{hamana01}.  This  method produced 348 slices in the
light cone, spanning the redshift  range [0,1].  Due to the large size
of the  simulated volume, the  effect of periodic replications  of the
computational box are minimized.  Each slice was then transformed into
a full-sky Healpix map ($nside=4096$) of the average overdensity using
a  simple ``Nearest Grid  Point'' (NGP)  mass projection  scheme.  The
density slices thus represented our physical model of the lens screens
used in  the ray-tracing procedure.  We  note that there  is no unique
procedure  for generating  a band-limited  harmonic  representation of
each slice of particles.  We chooe to use an NGP interpolation because
it is  a good compromise  between filtering and aliasing,  and remains
localised  in configuration  space.  More  sophisticated interpolation
schemes  have been  developed in  the  context of  either 3D  particle
distributions \citep{colombi}  or 2D continuous  fields \citep{basak},
which, however, remain impractical in significantly large simulations.

\begin{figure}
\resizebox*{!}{6cm}{\rotatebox{90}{ 
\includegraphics{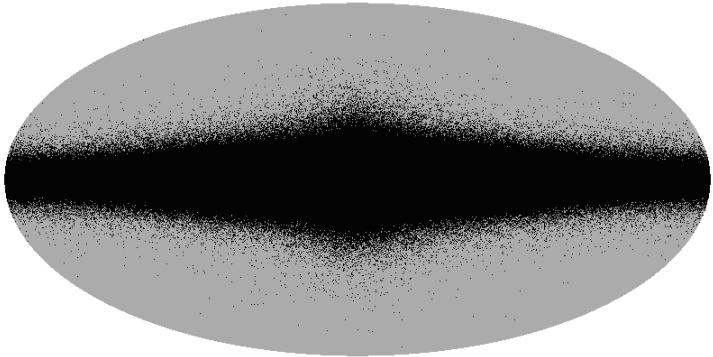}}}
\caption{Map  of the  cut-sky  used in  Sect.~\ref{stat}  to compute
  high-order moments.  }
\label{cutsky}
\end{figure}

After  an interpolation kernel  has been  chosen, all  fields (lensing
potentials  and  displacement  fields)   are  computed  from  the  NGP
interpolation mass slices at  each redshift using a spherical harmonic
decomposition.  The resampling of  the displacement fields outside the
pixel centres (as required in  a multi-lens method) is completed using
a   local  linear-interpolation   scheme   (using  covariant,   second
derivatives of  the potential); this  last interpolation has  the same
spectral behavior  (and thus the  same aliasing contamination)  as the
NGP-interpolated mass  slices, and we therefore  do not need  to use a
higher-order resampling scheme (since the calculation of the potential
requires two sets of integration over the mass distribution, while the
interpolation of the displacement  field corresponds to a second-order
derivative).  We provide  more details in Appendix~\ref{appendix} (see
\cite{jain00},  \cite{hamana01}   and  \cite{vale04}  for  alternative
approaches).   We assumed that  the background  galaxies are  within a
single  source   plane  located   at  redshift  $z_s=1$.    The  final
convergence  map  was  computed  using our  multiple-lens  ray-tracing
scheme, for which spherical geometry  precludes the use of small angle
approximations (as  in \cite{das})  especially in the  neighborhood of
the poles; full rotation matrices for each light ray must therefore be
computed from the displacement fields at each redshift.

The resulting full-sky Healpix map with a pixel size of $\Delta \theta
\simeq  0.74$  arcmin is  shown  in  Fig.~\ref{fullsky}, with  small
inserts     to     highlight     the     large     dynamical     range
achieved\footnote{Higher   resolution    images   are   available   at
  \texttt{http://www.projet-horizon.fr}.}.   The  particle shot  noise
corresponding to our 70 billion particle run has a small impact on the
map.  As shown in  Fig.~\ref{powerspectra}, the particle shot noise is
well below the expected  instrumental noise, and even sufficiently low
to be ignored in the spectral analysis of the signal.

\section{High--order moments and realistic sky cut}
\label{stat}
\begin{figure}
\resizebox*{!}{9cm}{\includegraphics{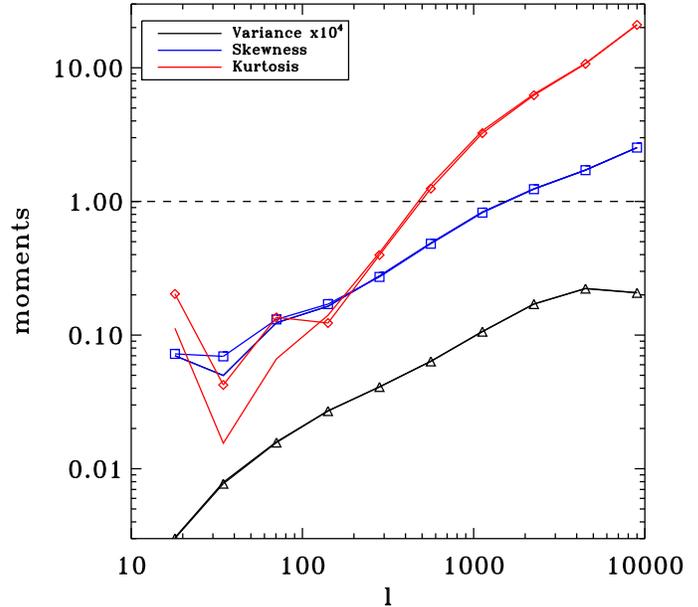}}
\caption{Moments  of the  convergence  as a  function  of the  average
  multipole moment on each  wavelet scale. The variance, skewness, and
  kurtosis   are    shown   as    black,   blue,   and    red   lines,
  respectively. Solid lines with  error bars corresponds ro a full-sky
  analysis, while dotted lines correspond to our cut-sky analysis. }
\label{statistics}
\end{figure}

\begin{figure}
\resizebox*{!}{9cm}{\includegraphics{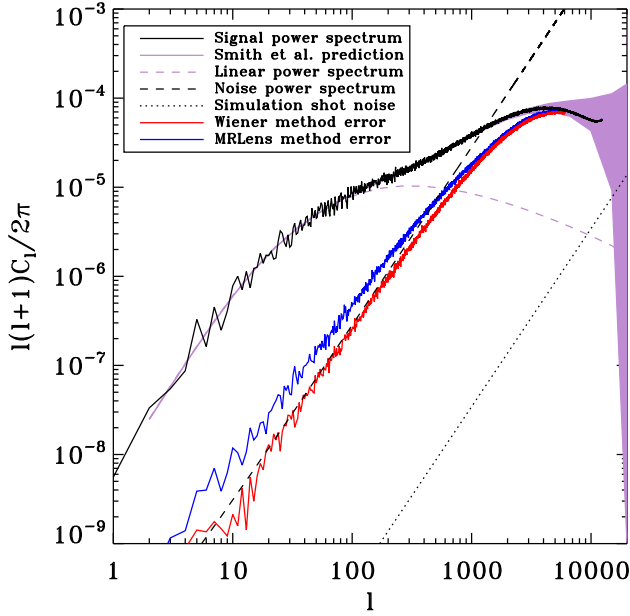}}
\caption{Angular  power  spectrum  of  the simulated  convergence  map
  (black  solid  line),  compared to  a  fit  based  on the  Smith  et
  al.  (2001) analytical model  with error  bars corresponding  to our
  noise model  (pink area). Also  shown is the prediction  from linear
  theory (pink dashed  curve). The noise power spectrum  is plotted as
  the dashed black  line.  The green solid line  is the power spectrum
  of the error  map obtained with the Wiener  filter method, while the
  blue solid line are that for the MRLens method.}
\label{powerspectra}
\end{figure}

\begin{figure*}
\centering
\vfill
\resizebox*{!}{5cm}{\includegraphics{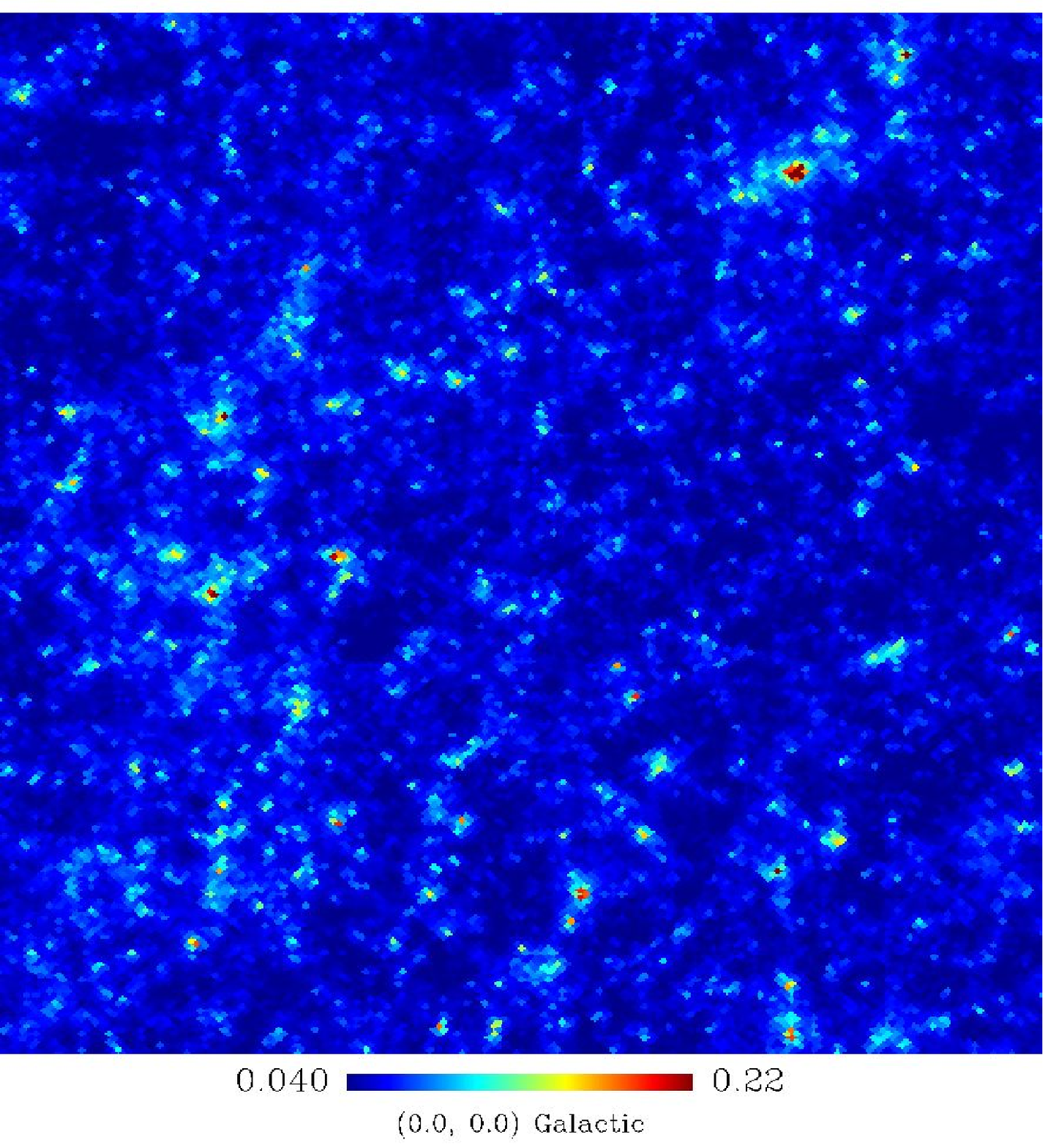}}
\resizebox*{!}{5cm}{\includegraphics{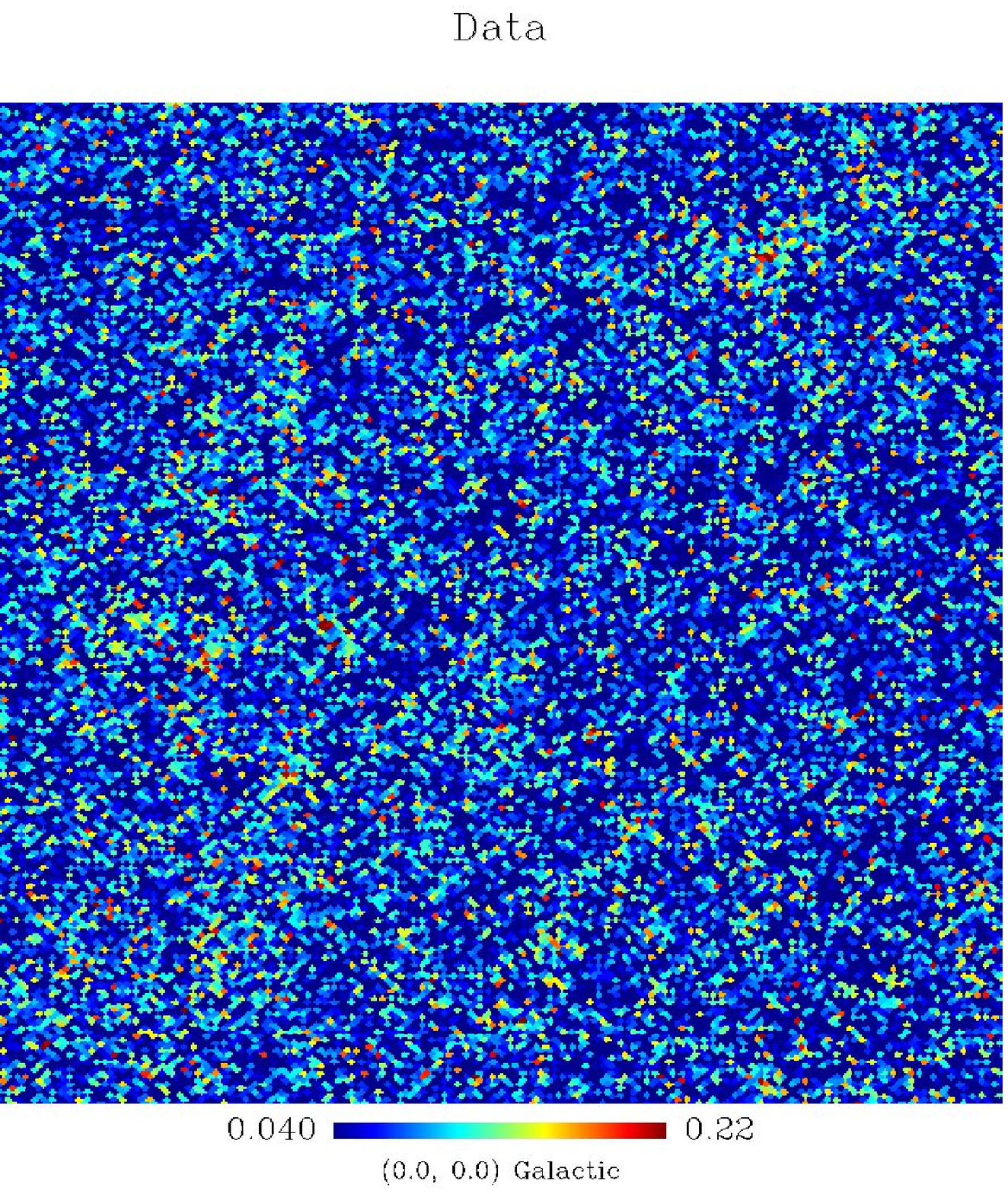}}
\resizebox*{!}{5cm}{\includegraphics{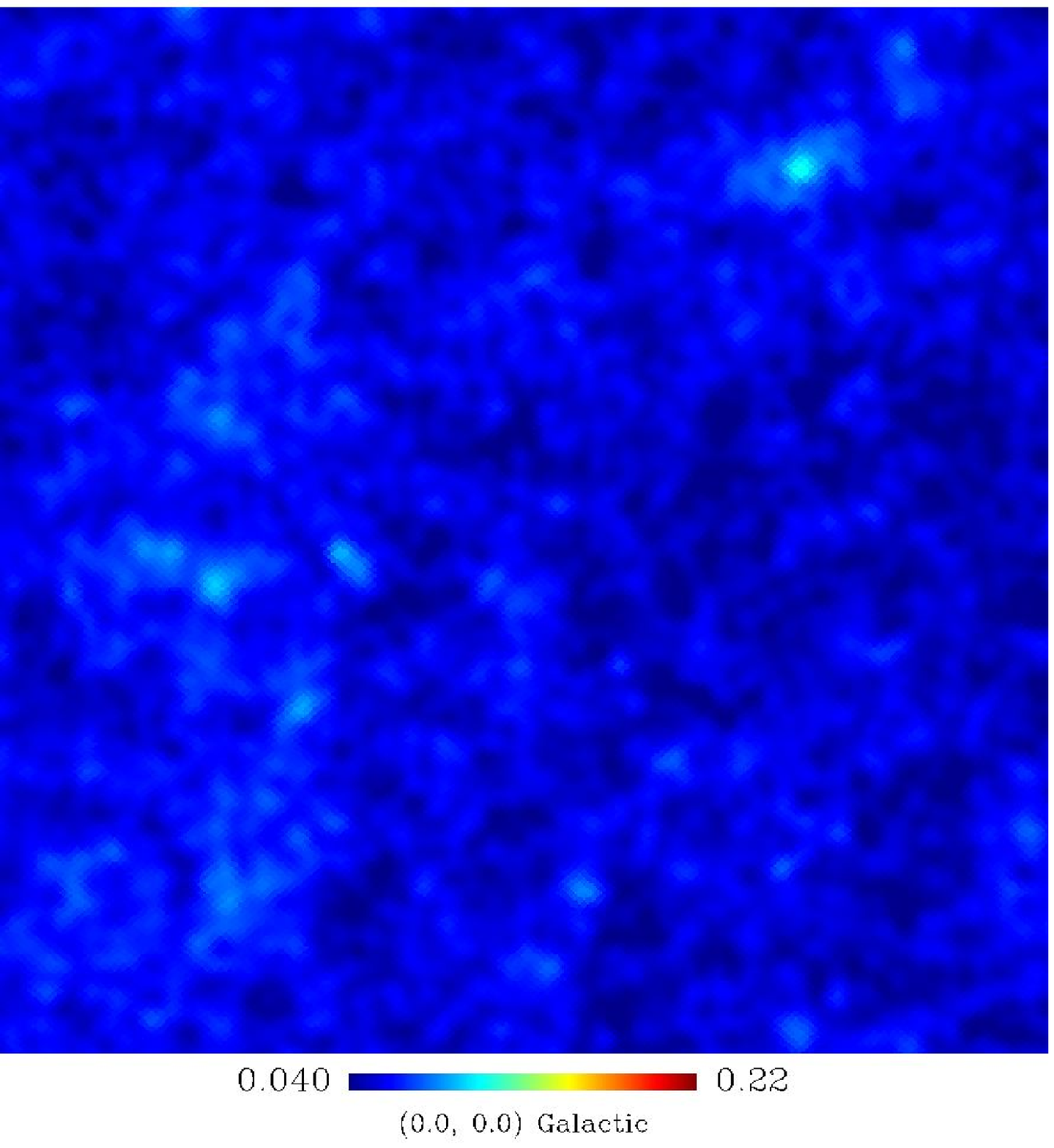}}
\resizebox*{!}{5cm}{\includegraphics{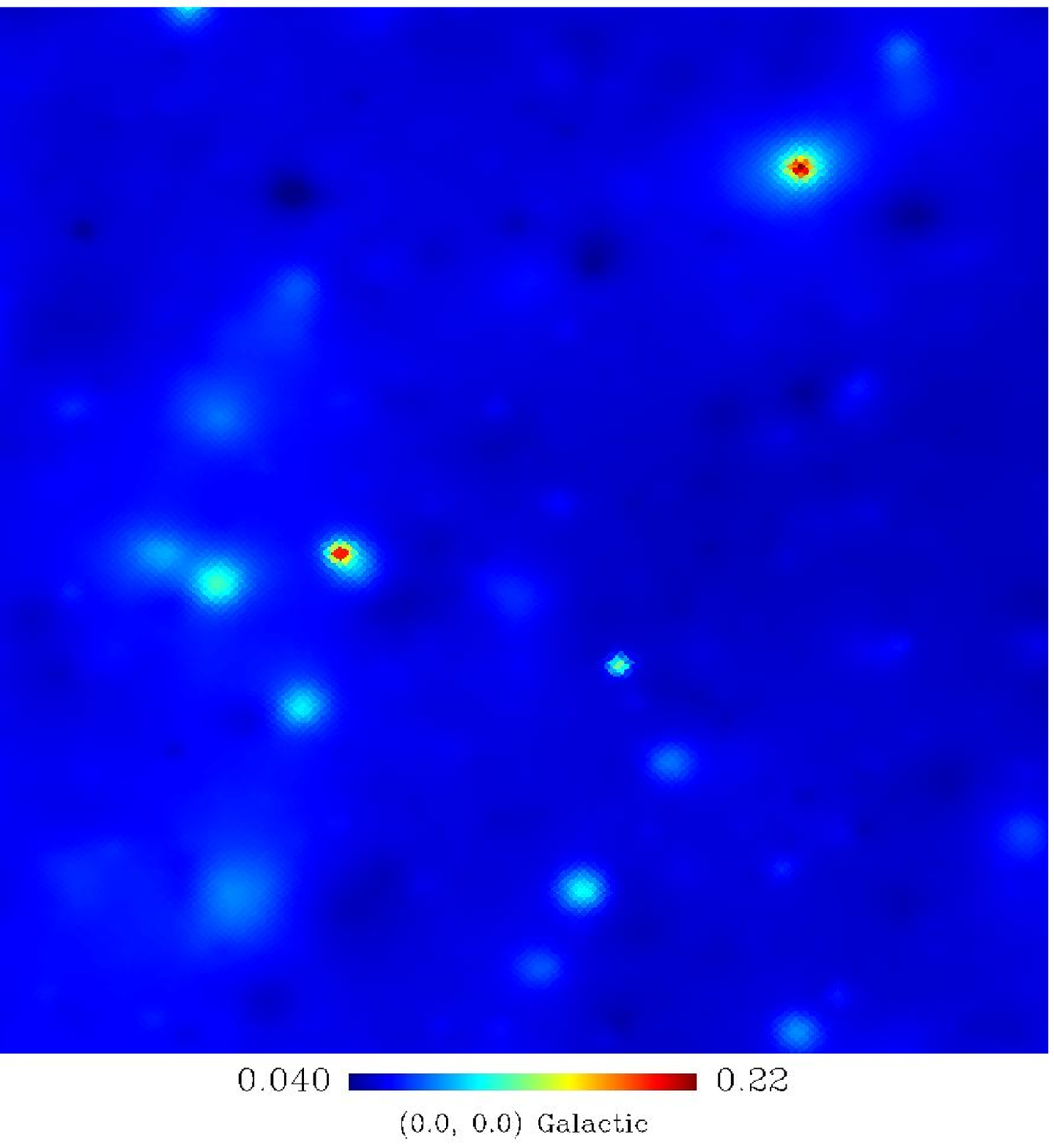}}
\vskip -0.5cm
\resizebox*{!}{5cm}{\includegraphics{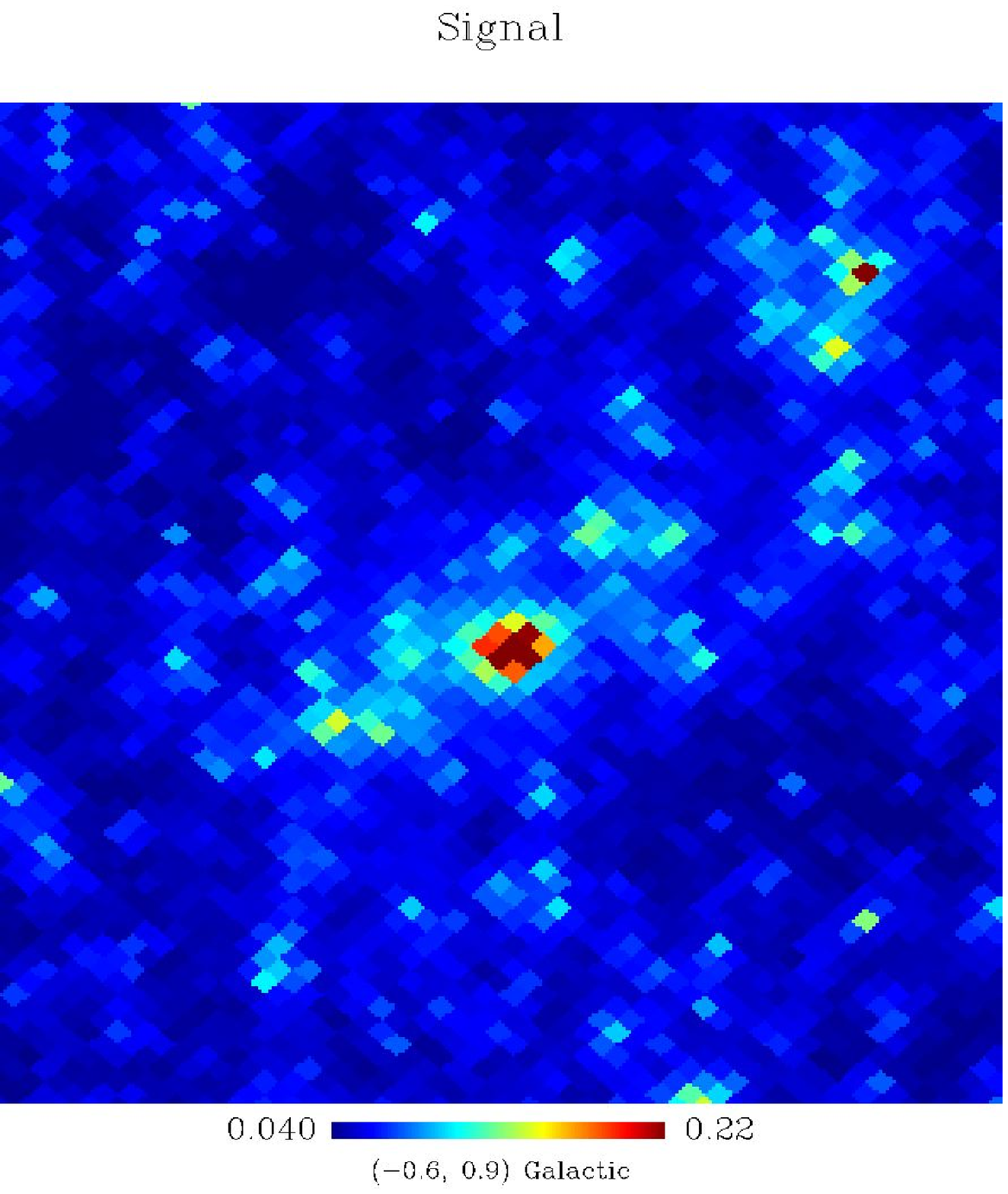}}
\resizebox*{!}{5cm}{\includegraphics{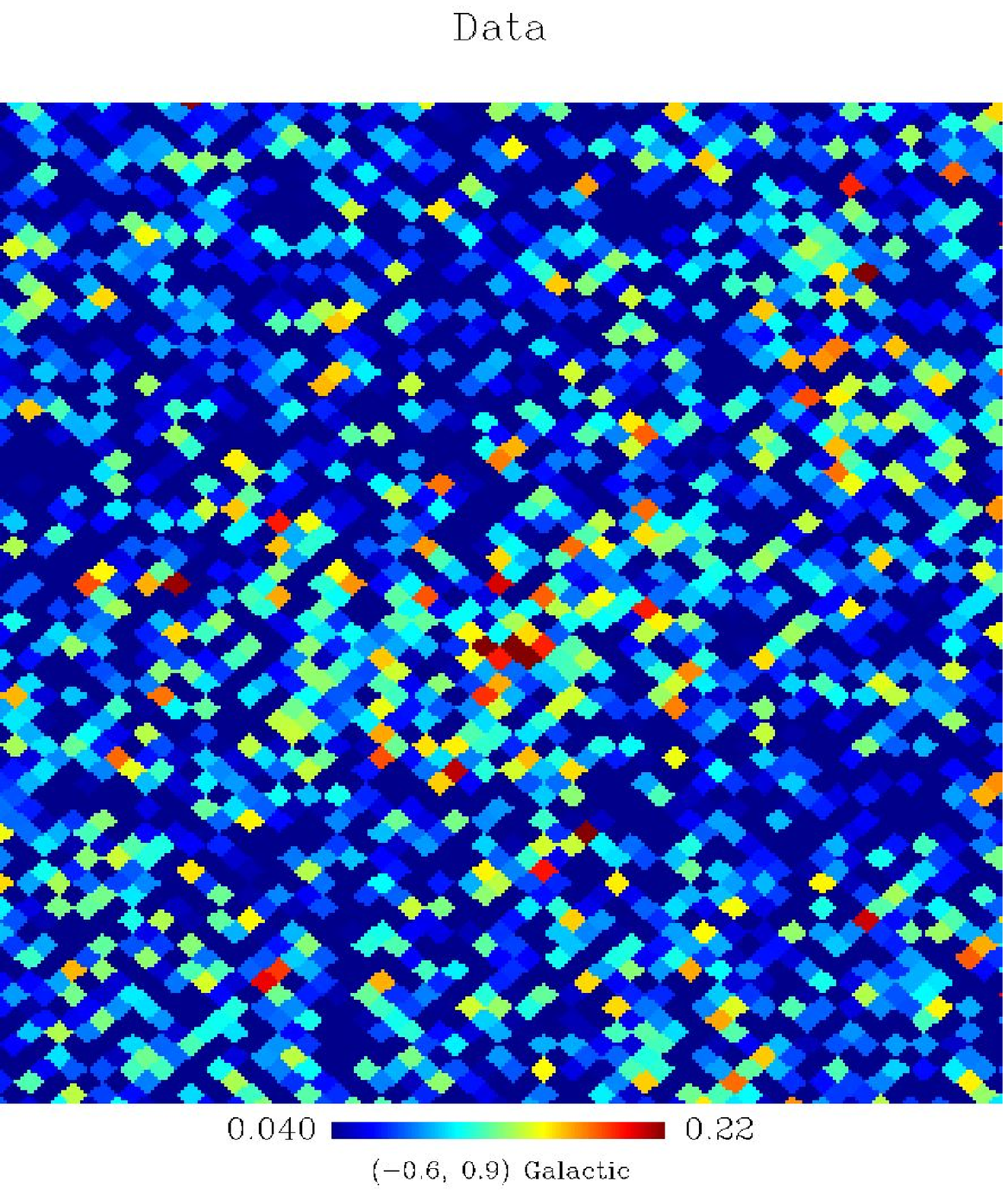}}
\resizebox*{!}{5cm}{\includegraphics{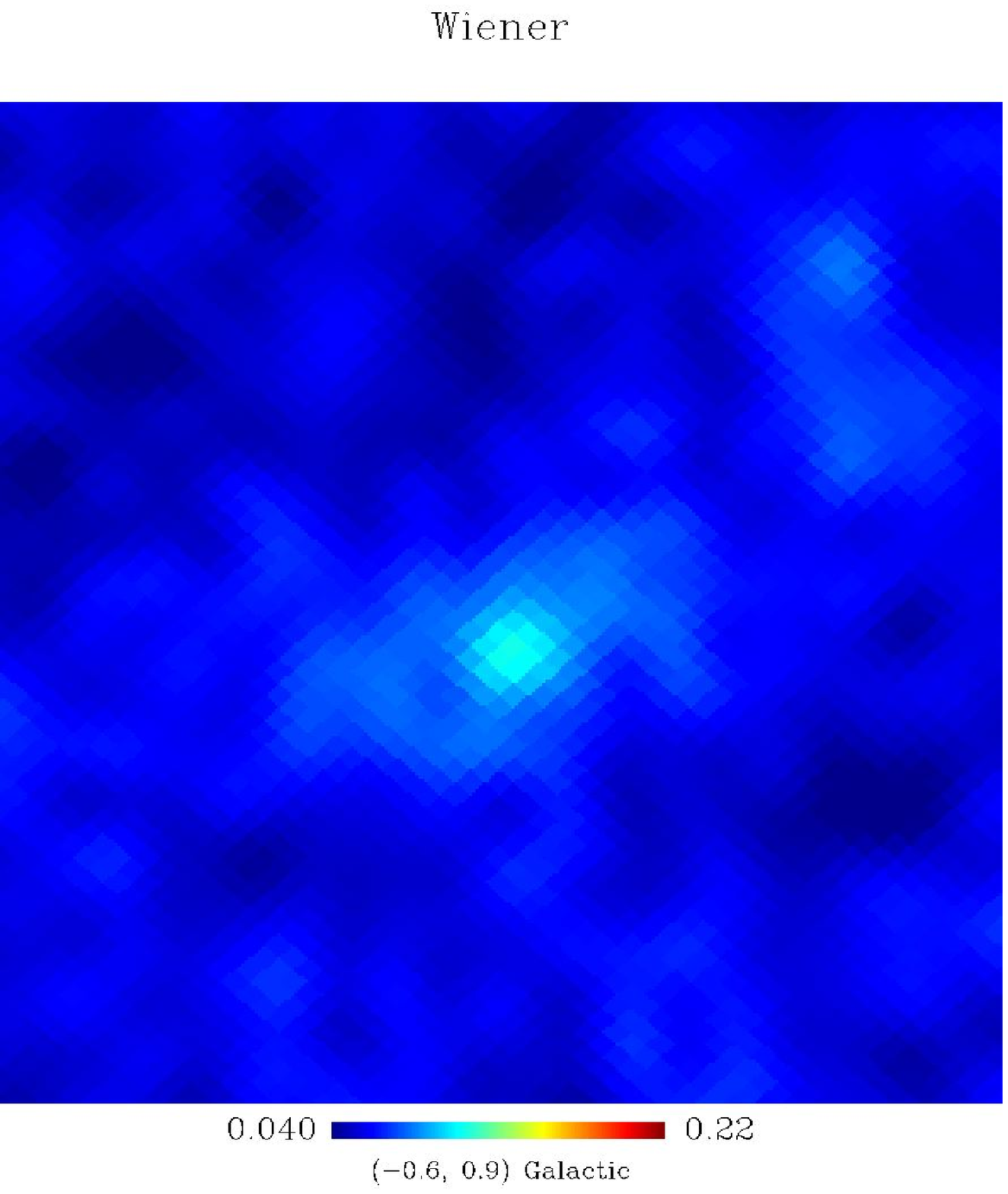}}
\resizebox*{!}{5cm}{\includegraphics{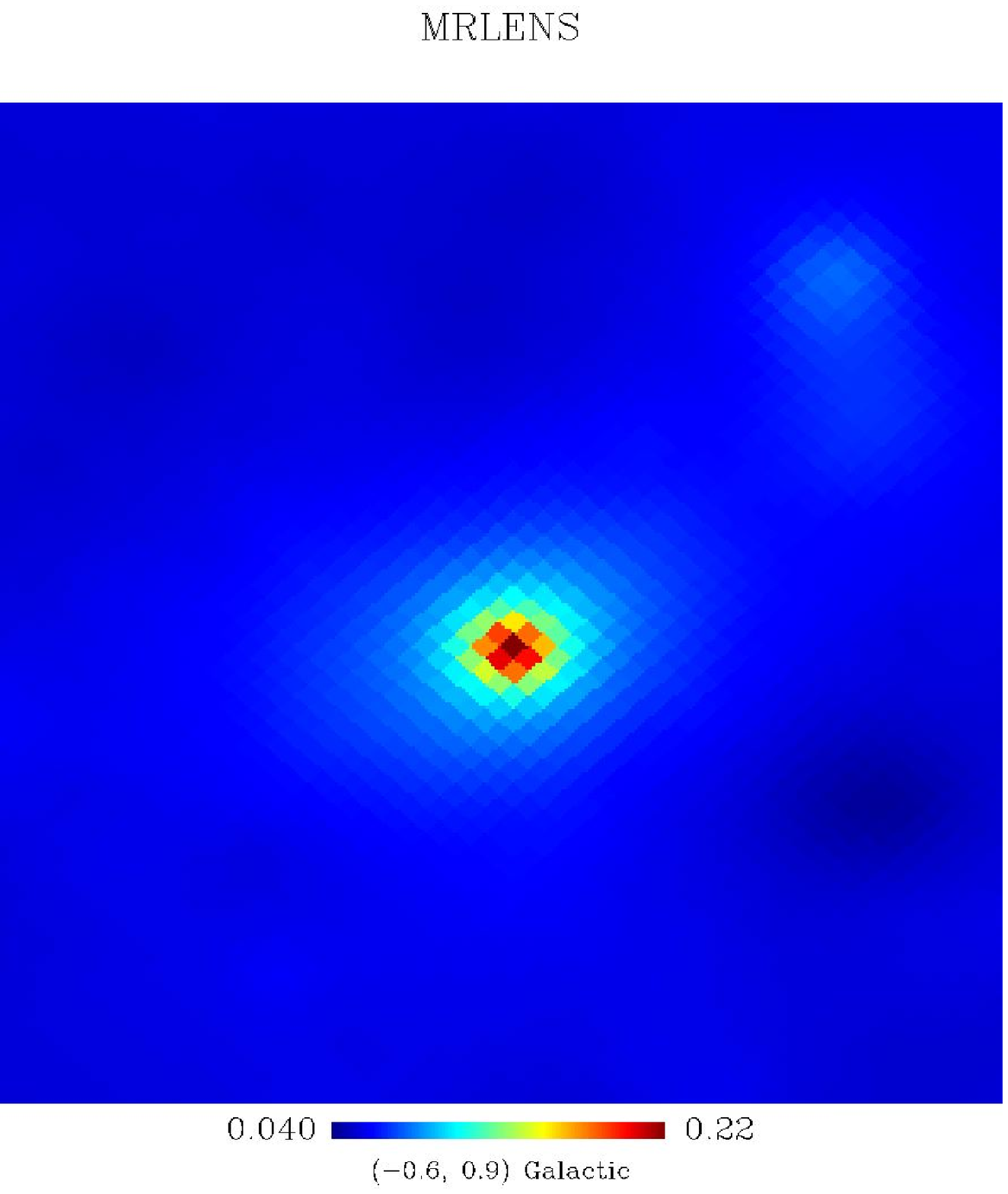}}
\vskip -0.1cm
\caption{Reconstruction  of  convergence  maps  with our  2  filtering
techniques.   The top  panels show  the $2.5^o\times2.5^o$  square map
corresponding  to first  zoom sequence  of  Fig.~\ref{fullsky}.  The
bottom panels are  subset of the corresponding top  images with linear
size 45$^{'}$.  From  left to right, we show  the original signal, the
noisy  image, the  Wiener-filtered image  and the  the MRLens-filtered
image.}
\label{filtering}
\end{figure*}

In Fig.~\ref{fullsky}, the signal appears as a typical Gaussian random
field on large scales, similar  to the Cosmic Microwave Background map
seen by  the WMAP satellite  \citep{spergel07}.  On small  scales, the
signal is  clearly dominated by clumpy structures  (dark matter halos)
and   is  therefore   highly  non-Gaussian.    To   characterize  this
quantitatively, we performed a  wavelet decomposition of our map using
the   Undecimated   Isotropic   Wavelet   Transform  on   the   sphere
\citep{moudden06}, and,  for each wavelet scale, we  have computed its
second-,  third- and  fourth-order  moment.  We  used  11 scales  with
central multipole values of $\ell_0=9000$, 4500, 2250, 1125, 562, 282,
141, 71,  35, 18.  For  each of these  maps, we computed  the variance
$\sigma^2  =  \langle\kappa^2\rangle$, the  normalized  skewness $S  =
\langle\kappa^3\rangle /  \sigma^3$, and the normalized  kurtosis $K =
\langle\kappa^4\rangle   /   \sigma^4$.    Results  are   plotted   in
Fig.~\ref{statistics} as  solid lines  of various colors.   Error bars
were  estimated  approximately by  computing  each  moment  on the  12
Healpix base  pixels independently and evaluating the  variance in the
12 results.   A more appropriate  strategy would have been  to perform
several,  independent, 70  billion particle  runs, which  is currently
impossible for us  to do.  We can see that the  variance in the signal
steadily increases for higher  and higher multipoles, and saturates at
a  fraction of $10^{-4}$,  corresponding to  the value  predicted from
nonlinear gravitational clustering for  $\ell \ge 6000$.  The variance
for each wavelet  plane can be considered to be  a band power estimate
of   the  angular   power   spectrum,  as   can   be  verified   using
Fig.~\ref{powerspectra}. In the same  figure, we have also plotted for
comparison  the {\it linear}  power spectrum,  to highlight  the scale
below which nonlinear  clustering contributes significantly, i.e., for
$\ell > 750$  or equivalently $\theta < 15^{'}$,  as first pointed out
by \cite{jain97}.  Skewness and kurtosis are more direct estimators of
the signal non--Gaussianity.  Departures from Gaussianity occur around
$\ell \simeq 750-1500$, where both statistics cross unity.  Due to the
large dynamical  range of  the Horizon simulation,  we computed  a map
spanning two decades in angular  scales in the linear, Gaussian regime
and  two additional  decades  in angular  scales  into the  nonlinear,
non--Gaussian regime.

It  is clear  from  Fig.~\ref{statistics} that  at  small $\ell$,  the
skewness and the  kurtosis of the map are  strongly affected by cosmic
variance.  The statistics of  the convergence field cannot be measured
in practice  over the  whole sky  because of sky  cuts imposed  by the
presence of saturated  stars and by absorption in  the Galactic plane.
We  estimated the  impact  of this  sky  cut on  the  accuracy of  our
multi-resolution statistical analysis. We computed the expected number
of bright stars that would  saturate CCD cameras typically employed in
wide-field  survey  (B-magnitude$< 20$).   We  then  removed from  our
analysis each pixel contaminated by  at least 3 bright stars, based on
a  random  Poissonian  realization  of  bright  stars  in  our  Galaxy
(according to  the model  presented in \cite{bahcall80},  Appendix B).
We obtain a  mask with 40\% of the  sky removed, corresponding roughly
to a $\pm 20^o$ cut around the Galactic plane (see Fig.~\ref{cutsky}).
The  resulting   statistics  are   overplotted  as  dotted   lines  in
Fig.~\ref{statistics}.  The transition  scale, for which the departure
from  Gaussianity is  significant,  can still  be estimated  reliabily
around $\ell \simeq 750-1500$.   We concluded that the cosmic variance
of the  cut sky  affects high--order moments  only below  $\ell \simeq
200$.

\section{Map--making using multi--resolution filtering}
\label{filter}

\begin{figure}
\resizebox*{!}{9cm}{\includegraphics{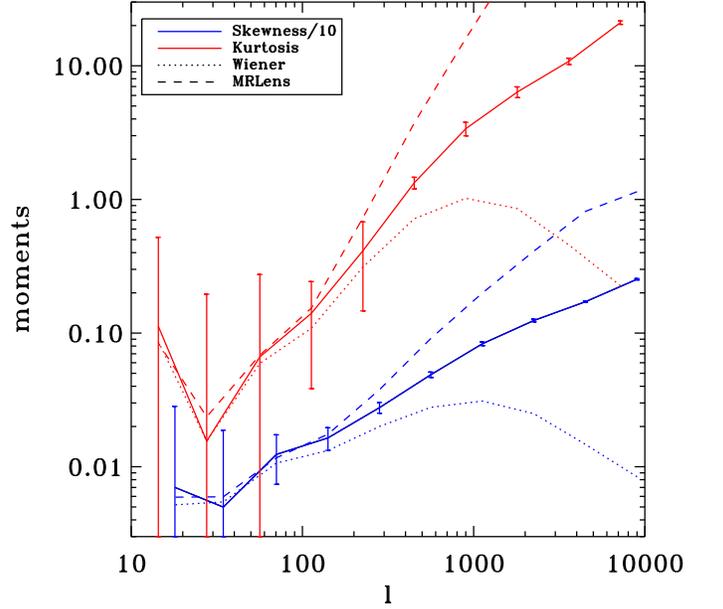}}
\caption{Skewness  (blue  lines)  and  kurtosis (red  lines)  for  the
  original convergence map (solid  lines with error bars), compared to
  the  same high-order  statistics  for the  Wiener reconstructed  map
  (dotted lines) and the MRLens reconstructed map (dashed lines). }
\label{merit}
\end{figure}

The full-sky simulated convergence maps described above can be used to
analyze and compare de-noising (or map--making) methods on the sphere.
For  this purpose, we  considered a  purely white  instrumental noise,
typical  of the  next generation  all--sky  surveys, and  a root  mean
square  per  pixel  of  area  $A_{p}$  given  by  $\sigma_N  =  0.3  /
\sqrt{n_{gal}A_p}$   for    $n_{gal}=40$   background   galaxies   per
arcmin$^2$.  Recovering  the most accurate convergence  map from noisy
data will be an important  step in future surveys.  This reconstructed
map can be used to construct a mass selected halo catalog, measure its
statistical properties  and constrain cosmological  parameters, and be
compared directly  with other  cluster catalogues compiled  with other
techniques (X-ray, galaxy counts or SZ).  We restrict ourselves to the
{\it full-sky}  denoising of  a convergence map  already reconstructed
from  the shear  derived from  galaxy ellipticities.   In  the present
work, we do not address filtering  in the presence of a cut--sky, such
as  the one shown  in Fig.~\ref{cutsky}.   Promising methods  based on
``impainting''    have   been   developed    in   the    CMB   context
\citep{Abrial08}, and  also weak-lensing applications \citep{Pires08};
these replace missing  data with an artificial signal  and allow us to
optimize the results we obtained with filtering methods for a full-sky
analysis.

A  straightforward filtering  method is  the Wiener  filtering scheme,
which  is optimal  for  Gaussian  random fields,  and  is expected  to
operate here  effectively on large  scales.  Defining $S_\ell$  as the
power spectrum  of the input signal  (see Fig.~\ref{powerspectra}) and
$N_\ell$  the  power  spectrum  of  the noise,  this  method  involves
convolving   the  noisy   map  by   the  Wiener   filter   defined  as
$W_\ell=S_\ell  /  (S_\ell +  N_\ell)$.   The  results  of the  Wiener
filtering approach are  shown in Fig.~\ref{filtering}.  Comparing with
the  input signal  map, we  conclude that,  although the  agreement is
satisfactory on large scales, the  dense clumps clearly visible in the
image  are  poorly recovered  because  they  have  been convolved  too
significantly.

\begin{figure}
\resizebox*{!}{9cm}{\includegraphics{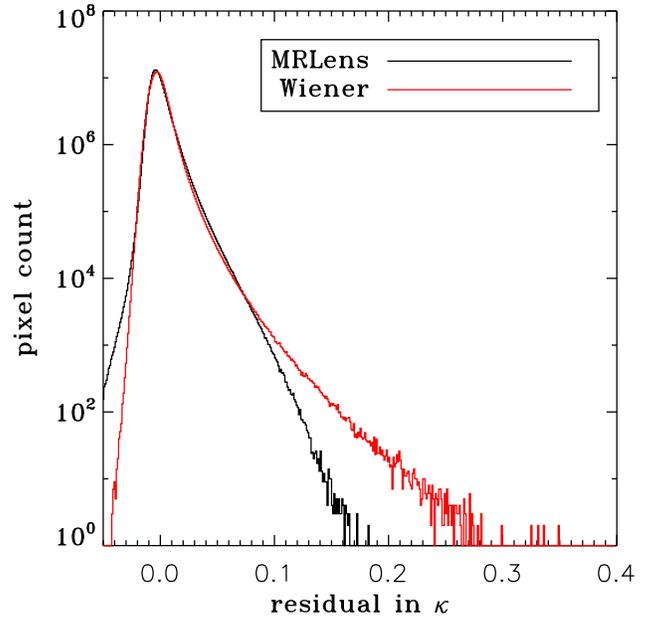}}
\caption{Histogram of the residual maps for Wiener and MRLens filtering.
  }
\label{pdfres}
\end{figure}

A  dedicated weak-lensing  wavelet-restoration method,  called MRLens,
has been  developed \citep{pires06}.   It can be  considered to  be an
extension of the Maximum  Entropy Method (MEM) that provides different
types  of  information.  In  MRLens,  the  entropy  constraint is  not
applied  to  the  pixels  of  the solution,  but  rather  its  wavelet
coefficients.  This  allows us to  take into account  more efficiently
the  multi-scale behavior  of the  information.  MRLens  was, however,
designed for weak-lensing maps of smaller surface area on the sky, for
which the  non--Gaussian signal is stronger. MRLens  was extended here
to the sphere by considering independently each of the 12 Healpix base
pixels covering the sphere as  12 independent Cartesian maps, on which
we applied the MRLens algorithm of \cite{pires06}.  Full-sky denoising
performed with  MRLens is shown in  Fig.~\ref{filtering}.  It performs
more efficiently than  the Wiener methods on small  scales, with dense
clumps more  accurately estimated, but less efficient  than the Wiener
method on large scale when  recovering low frequency waves in the map.
We also computed the angular power spectrum of the error map (see
Fig.~\ref{powerspectra}) in both cases  (Wiener and MRLens). We can see
that     Wiener    filtering     outperforms    MRLens     on    large
scales. Interestingly, the  MRLens errors decrease significantly above
the transition  scale we identified in  the last  section around
$\ell \simeq 1000$ (see Fig.~\ref{powerspectra}).

To compare both methods  more quantitatively, we computed the skewness
and  kurtosis  of  both  reconstructed  maps.  Results  are  shown  in
Fig.~\ref{merit}. We note that  using map-making algorithms to recover
the skewness and kurtosis of the true signal is not at all the optimal
strategy: maximum likelihood estimators are more appropriate.  We used
high-order statistics here only to compare the relative merits of each
method.  It is striking to observe in Fig.~\ref{merit} that the Wiener
reconstructed map strongly underestimate the skewness and the kurtosis
at small scale.  This confirms quantitatively what was already visible
in  the maps  (Fig.~\ref{filtering}),  namely that  the Wiener  method
strongly suppresses high  peaks in the map, affecting  the tail of the
probability  distribution function.   On  the other  hand, the  MRLens
reconstructed  map  has  a  significantly {\it  higher}  skewness  and
kurtosis  than the original  map: this  wavelet--based method  is only
efficient  in  recovering high  peaks  in  the  signal, affecting  the
reconstructed probability density function in the opposite direction.

We now use the Probability Density Function (PDF) of the residual maps
to compare each method (see Fig.~\ref{pdfres}).  We confirm our visual
impression  from   Fig.~\ref{filtering}  that  MRLens   performs  more
efficiently than the Wiener method in recovering the high convergence,
nonlinear features  in the  map.  The positive  high residual  tail is
reduced  significantly by  MRLens, as  well as  the dozen  of strongly
outlying pixels in the Wiener filterer map around $\kappa \simeq 0.35$
(see  Fig.~\ref{pdfres}). MRLens, however,  performs poorly  for small
values of  the convergence ($\kappa  \simeq \pm 0.05$), for  which the
PDF  is well  approximated by  a  Gaussian, an  optimal situation  for
Wiener filtering.

The present  analysis, based on using  both the power  spectrum of the
residual  maps  and the  high--order  moments  of  the recovered  map,
strongly suggests that  new methods should be developed  using an {\it
  hybrid}, multi-resolution formulation; for instance, using spherical
harmonics on  large scales,  while utilizing wavelets  coefficients on
small  scales.  The  methodology of  this combined  approach  could be
based   on   the   idea    of   Combined   Filtering   introduced   by
\cite{moudden06}.

\section{Conclusion}
\label{conclusion}
Using the 70  billion particles of the Horizon  N--body Simulation, we
have computed for the first  time a realistic full-sky convergence map
with   a   pixel   resolution   of   $\Delta   \theta   \simeq   0.74$
arcmin$^{2}$.   We    have   analyzed   the    resulting   map   using
multi-resolution  statistics (variance,  skewness,  and kurtosis)  and
angular power-spectrum  analysis.  We  have shown that  this simulated
map spans 4 decades of useful  signal in angular scale, with 2 decades
within  the  linear, Gaussian  regime  and  2  decades well  into  the
nonlinear, non--Gaussian regime.  We have shown that, when considering
a realistic sky cut, we  can reliabily estimate high--order moments of
the map above  $\ell \simeq 200$.  Using even  higher resolution maps,
angular scales smaller than $\theta \simeq 1^{'}$ could be explored in
future works, although  the mass ditribution on these  scales might be
affected by  baryons physics \citep{jing06},  so that the  present map
might already cover all cosmologically relevant scales.

As a  first step  towards a realistic  map--making procedure,  we have
tested  two  de-noising schemes  on  a  simplified fictitious  dataset
derived from the full-sky map,  namely Wiener filtering and the MRLens
method  \citep{pires06}.   We have  shown  quantitatively that  Wiener
filtering is the most effective  method on large scales, although some
signal is lost  on small scales.  MRLens performs  more effectively on
small scales and recovers the dense clumps associated with dark matter
halos, but deals less accurately  with low frequency waves in the map.
Hence,  this work  demonstrates the  need for  hybrid multi-resolution
approach,   e.g.,  by  combining   spherical  harmonics   and  wavelet
coefficients.  The present analysis will be extended in future work to
map--making  algorithms  dealing  directly  with galaxy  shears.   The
simulated convergence  map may prove to  be an effective  tool for the
design of new map--making methods  and for the preparation of the next
generation  weak--lensing   surveys\footnote{The  convergence  map  is
  freely          available          for          download          at
  \texttt{http://www.projet-horizon.fr}.  }.

\begin{acknowledgement}
We  would  like  to  thank  Julien Devriendt,  Pierre  Ocvirk,  Arthur
Petitpierre and Philippe Lachamp  for their unvaluable help during the
course  of this  project. The  Horizon Simulation  presented  here was
supported by  the ``Centre de Calcul Recherche  et Technologie'' (CEA,
France) as a  ``Grand Challenge project''. This work  was supported by
the  Horizon Project.  Some  of the  results in  this paper  have been
derived using the HEALPix \citep{gorski05} package.
\end{acknowledgement}

\appendix

\section{Computing the convergence maps from simulations}

\label{appendix}

\noindent 
We  first  recall   how  to  compute  the  convergence   in  the  Born
approximation, and then present our new ray-tracing scheme.

\subsection{\noindent Born approximation}

\noindent 
We  start by  the  formula  relating the  convergence  to the  density
contrast:
\[
\kappa(\hat{n})=\frac{3}{2}\Omega_{m}\int_{0}^{z_{s}}\frac{dz}{E(z)}\frac{\mathcal{D}(z)\mathcal{D}(z,z_{s})}{\mathcal{D}(z_{s})}\frac{1}{a(z)}\delta(\frac{c}{H_{0}}\mathcal{D}(z)\hat{n},z) \,,
\]
which  is  valid  for  sources  at  a  single  redshift  $z_{s}$,  and
$\mathcal{D}(z)=\frac{H_{0}}{c}\chi(z)$    is    the    dimensionless,
comoving, radial coordinate  ($d\mathcal{D}=dz/E(z)$).  We now rewrite
this  formula  in a  form  that is  more  suited  to integration  over
redshift slices in a simulation:
\[
\kappa(\theta_{pix})\approx\frac{3}{2}\Omega_{m}\sum_{b}W_{b}\frac{H_{0}}{c}\int_{\Delta z_{b}}\frac{cdz}{H_{0}E(z)}\delta(\frac{c}{H_{0}}\mathcal{D}(z)\hat{n}_{pix},z)\,,
\]
where 
\[
W_{b}=\left(\int_{\Delta z_{b}}\frac{dz}{E(z)}\frac{\mathcal{D}(z)\mathcal{D}(z,z_{s})}{\mathcal{D}(z_{s})}\frac{1}{a(z)}\right)/\left(\int_{\Delta z_{b}}\frac{dz}{E(z)}\right)
\]
is a slice-related weight, and  the integral over the density contrast
reads
\begin{eqnarray*}
I & = & \int_{\Delta z_{b}}\frac{cdz}{H_{0}E(z)}\delta(\frac{c}{H_{0}}\mathcal{D}(z)\hat{n}_{pix},z)\,,\\
& = & \int_{\Delta\chi_{b}}d\chi\delta(\chi\hat{n}_{pix},\chi)\,,\\
& \approx & \frac{V({\rm sim})}{N_{part}({\rm sim})}\left(\frac{N_{part}(\theta_{pix},z_{b})}{S_{pix}(z_{b})}-\Delta\chi_{b}\right)\,,
\end{eqnarray*}
where 
\[
S_{pix}(z_{b})=\frac{4\pi}{N_{pix}}\frac{c^{2}}{H_{0}^{2}}\mathcal{D}^{2}(z_{b})
\]
is the comoving surface of  the spherical pixel. Interpreting all together,
we obtain the  following formula for the convergence  map (omitting the
$\Delta\chi_{b}$ term that corresponds to a constant term):
\begin{equation}
\kappa(\theta_{pix})=\frac{3 \Omega_{m}}{2}\frac{N_{pix}}{4\pi}\left(\frac{H_{0}}{c}\right)^{3}\frac{V({\rm sim})}{N_{part}({\rm sim})}\sum_{b}W_{b}\frac{N_{part}(\theta_{pix},z_{b})}{\mathcal{D}^{2}(z_{b})}\,.
\label{eq:kappaborn}
\end{equation}
\noindent This is the equation used to derive the convergence map in
the Born approximation.

\subsection{Ray--tracig using multiple planes}

We  discuss  here   the  formulae  needed  for  the  multi-plane
computations, where we consider the lensing by a number of thin lenses
located at $\{ z_{b}\}$.  We define
\[
\kappa_{fac}=\frac{3}{2}\Omega_{m}\frac{N_{pix}}{4\pi}\left(\frac{H_{0}}{c}\right)^{3}\frac{V({\rm sim})}{N_{part}({\rm sim})}\,,
\]
and                                                    
\begin{equation}
  \zeta(z_{b},\theta)=\kappa_{fac}\omega(z_{b})\frac{N_{part}(\theta,z_{b})}{\mathcal{D}^{2}(z_{b})}\,,\label{eq:zetaloc}
\end{equation}
with                
\[                \omega(z_{b})=\left(\int_{\Delta
  z_{b}}\frac{dz}{E(z)}\frac{\mathcal{D}(z)}{a(z)}\right)/\left(\int_{\Delta
  z_{b}}\frac{dz}{E(z)}\right)\,.
\]  
To follow the  light rays, we are interested  in computing the angular
displacement field  for each ray  $i$ due to  a slice at  $z_{b}$.  We
then define
\begin{equation}
\alpha_{i}^{b}=\left(-2\nabla\Delta^{-1}\left(\zeta(z_{b})\right)\right)
(\theta_{i})\,,\label{eq:deflection}
\end{equation}
where the gradient and  Laplacien are computed using angular covariant
derivatives  on  the (unit)  sphere,  and  $\theta_{i}$ is the  current
direction of  light ray $i$  when it is incident on  the slice $b$. Now,  
we
start from light  rays that are back-propagated from the  observer at z=0
towards the  source (here at  z=1). We denote  by $\{\theta^{1}\}$
the location of the Healpix  centres, which corresponds to the initial
directions of  the back-propagated  rays emanating from  the observer.
The  tangent  vectors  to each  light  ray  will  be modified  by  the
deflection   field  at   each   lens  plane,   defined  by   Eq.~
\ref{eq:deflection}.  Then, computing the  displacement of the rays at
slice $b$ reads
\[
\vec{\alpha}_{i}^{b}=
\left(-2\nabla\Delta^{-1}(\zeta(z_{b})\right)(\theta_{i}^{b})\,.
\]
We  then  update  the  direction  $\vec{\beta}_{i}^{b}$  of  the  rays
according to the following rotation, $\cal R$:
\begin{equation}
  \vec{\beta}_{i}^{b}={\cal R}(\vec{n}_{i}^{b}\times\vec{\alpha}_{i}^{b},\|\vec{\alpha}_{i}^{b}\|)\vec{\beta}_{i}^{b-1}\,.
  \label{eq:rotateray}
\end{equation}
where $\vec{\beta}_{i}^{0}=\vec{n}_{i}^{1}$ (light  rays emanate from
the observer,   thus  in a direction perpendicular    to   the   first   slice),   and
$\vec{n}_{i}^{b}$is the vector normal to slice $b$ at the intersection
of  light-ray $i$  on slice  $b$.  Equation~(\ref{eq:rotateray})  can be
simplified by noting that $\vec{\alpha}$ is expressed naturally in the
local $(\vec{e}_{\theta,}\vec{e}_{\phi})$  basis of the  tangent plane
at position $(\theta,\phi)$:
\[
\vec{\alpha}=\|\vec{\alpha}\|(\cos\delta\vec{e}_{\theta}
+\sin\delta\vec{e}_{\phi})\,.
\] 
After calculating the new value of  $\vec{\beta}$, one needs to
compute the intersection  of the light rays with  the next shell.
We call $\vec{x}_{i}^{b}$ the Cartesian position of the intersection of
light ray $i$ with slice $b$, then the next intersection will be given
by the solution for $\lambda$ of the system:
\begin{eqnarray*}        
\vec{x}_{i}^{b+1}        &       =        &
\vec{x}_{i}^{b}+\lambda\vec{\beta}_{i}^{b}\\ \lambda^{2}+2\lambda(\vec{x}_{i}^{b}\cdot\vec{\beta}_{i}^{b})+R_{b}^{2}-R_{b+1}^{2}
&  = & 0\,,\quad \lambda>0\,,
\end{eqnarray*}  
assuming that  $\vec{\beta}$ remains strictly unitary,  and $R_{b}$ is
the comoving radius of slice $b$.
 Once $\vec{x}_{i}^{b+1}$ is known, it
is   easy  to  compute   the  new   $\theta_{i}^{b+1}$ positions.   The
contributions    to   $\kappa$    are   then    calculated   following
Eq.~\ref{eq:kappaborn},  but where  the  slice contributions  are
interpolated at the displaced positions:
\[ 
\kappa(\theta_{i=pix})=\frac{3}{2}\Omega_{m}\frac{N_{pix}}{4\pi}
\left(\frac{H_{0}}{c}\right)^{3}\frac{V({\rm sim})}{N_{part}({\rm sim})}
\sum_{b}W_{b}\frac{N_{part}(\theta_{i}^{b},z_{b})}{\mathcal{D}^{2}(z_{b})}\,.
\]
We note that $\theta_i^b$ may fall into different pixels as a function of the slice $b$.
\bibliographystyle{aa} \bibliography{biblio}

\end{document}